\documentclass[useAMS,usenatbib]{aa}  

\usepackage{graphicx}
\usepackage{txfonts}
\usepackage{graphicx} 
\usepackage{times}
\usepackage{epstopdf}
\usepackage{amsfonts}
\usepackage{amsmath}
\usepackage{amsbsy}
\usepackage{bm}
\usepackage{url}
\usepackage{microtype}
\usepackage{rotating}
\usepackage{sidecap}
\usepackage{longtable}
\usepackage{lscape}
\usepackage{rotating}
 \usepackage{array,multirow,graphicx}
\usepackage[export]{adjustbox}
\usepackage{amssymb,xcolor}
\usepackage{colortbl}
\usepackage{booktabs}
\usepackage{url}
\usepackage[utf8]{inputenc}

\newcommand{\solarmass}{\(\textup{M}_\odot\)}

\def\apjs{ApJ}

\def\aap{A\&A}
\def\apjl{ApJ}

\begin{document}

   \title{MeerKAT discovers a jet-driven bow shock near GRS 1915+105}

   \subtitle{How an invisible large-scale jet sculpts a microquasar’s environment}
 \titlerunning{Jet-driven bow shock near GRS1915+105}
   \author{S.E. Motta
          \inst{1,2},
          P. Atri
          \inst{3,4},
          James H. Matthews
          \inst{2},
          Jakob van den Eijnden
          \inst{4},
          Rob P. Fender
          \inst{2},
          James C.A. Miller-Jones,
          \inst{6}
          Ian Heywood
          \inst{2,7,8}
          and Patrick Woudt
          \inst{5}}

    \authorrunning{Motta and Atri et al.}

   \institute{Istituto Nazionale di Astrofisica, Osservatorio Astronomico di Brera, via E.\,Bianchi 46, 23807 Merate (LC), Italy\\
              \email{sara.motta@inaf.it}
            \and 
              University of Oxford, Department of Physics, Astrophysics, Denys Wilkinson Building, Keble Road, OX1 3RH, Oxford, United Kingdom
            \and 
              ASTRON, Netherlands Institute for Radio Astronomy, Oude Hoogeveensedijk 4, 7991 PD Dwingeloo, The Netherlands\\
              \email{atri@astron.nl}
              \and 
              Anton Pannekoek Institute for Astronomy, University of Amsterdam, Postbus 94249, 1090 GE, Amsterdam, The Netherlands
              \and 
              Department of Astronomy, University of Cape Town, Private Bag X3, 7701 Rondebosch, South Africa
              \and 
              International Centre for Radio Astronomy Research, Curtin University, GPO Box U1987, Perth, WA 6845, Australia
              \and 
              Department of Physics and Electronics, Rhodes University, P.O. Box 94, Makhanda, 6140, South Africa
              \and 
              South African Radio Astronomy Observatory, 2 Fir Street, Observatory 7925, South Africa
             }

   \date{Received September 31st, 2024; accepted January 26th, 2025}

 
  \abstract
{Black holes, both supermassive and stellar-mass, impact the evolution of their surroundings on a large range of scales. While the role of supermassive black holes is well studied, the effects of stellar-mass black holes on their surroundings, particularly in inducing structures in the interstellar medium (ISM), remain under explored. }
{This study focuses on the black hole X-ray binary GRS 1915+105, renowned for its active jets, and the primary aim is to unveil and characterise the impact of GRS 1915+105 on its environment by identifying structures induced by jet-ISM interaction.}
{We observed GRS 1915+105 with MeerKAT for a total exposure time of 14~hr, and we obtained the deepest image of GRS 1915+105 to date. Using a previously proposed self-similar model for large-scale jets, we inferred the properties of both the jets and the ISM, providing insights into the jet-ISM interaction site.}
{Our observations revealed a bow shock structure near GRS 1915+105, likely induced by a jet interacting with the ISM and blowing an overpressured cavity in the medium. We constrained the ISM density to 100--160 particles\,cm$^{-3}$ while assuming a temperature range of 10$^4$--10$^6$\,K, which implies a bow shock expansion velocity of $20\,{\rm km\,s}^{-1}<\dot{L} <\,360\,{\rm km\,s}^{-1}$. We estimate that the jet responsible for the formation of the bow shock has an age between 0.09 and 0.22 Myr, and the time-averaged energy rate transferred by such jets into the ISM is constrained to 3.3$\times$10$^{37}$\,ergs s$^{-1}<$ Q$_{jet}^{a}<$ 1.5$\times$10$^{39}$\,ergs s$^{-1}$. }
{Our results confirm that in stellar-mass black holes, the energy dissipated through jets can be comparable to the accretion energy, and through the interaction of the jet with the ISM, such energy is transferred back to the environment. 
This feedback mechanism mirrors the powerful influence of supermassive black holes on their environments, underscoring the significant role a black hole's activity has in shaping its surroundings.
}

   \keywords{X-ray: binaries -- accretion, accretion disks -- relativistic processes -- black hole physics -- stars: black holes
            }
            
   \maketitle
%

\section{Introduction}


%


Accretion of matter onto celestial bodies is a universal phenomenon giving rise to the formation of jets and other types of outflows observed across various scales and scenarios, from proto-planetary disks to gamma-ray bursts and from stellar-mass to super-massive black holes (SMBHs). 
In the case of accreting stellar-mass black holes (BHs), this process reaches its extremes: These objects contrive to feed back to the environment via jets and winds a large fraction of the energy and matter they could potentially have swallowed, thereby actively impacting their environment rather than behaving only as sinks. 
The significance of this feedback extends across diverse scales, from SMBHs fuelling active galactic nuclei (AGNs) to stellar-mass BHs in X-ray binaries, which are considered small-scale analogues of SMBHs \citep[see e.g. ][]{Merloni2003, Falcke2004, Motta2017, Fernandez-Ontiveros2021}. The influence of SMBH activity is believed to play a crucial role in regulating galactic-scale activities, for example, by triggering/quenching star formation and affecting the chemical enrichment of the intergalactic medium as well as the overall evolution of galaxies and large structures \citep[][]{Magorrian1998, Croton2006}. Similarly, stellar-mass BHs impact their surroundings by reintroducing a fraction of the infalling gas and liberating gravitational energy, thus energising the interstellar medium (ISM) and increasing/decreasing its density and even injecting high energy particles into the ISM \citep{LHAASO2024, Marti-Devesa2024}. These phenomena induce interstellar turbulence that may stimulate local star formation and may even introduce seed magnetic fields into the ISM, potentially contributing to the average magnetic field observed in the Galaxy \citep[see e.g.][]{Heinz2008, Mirabel2015}.



Theoretical models describing the interaction between relativistic jets from BHs and their environment have postulated the emergence of a shock. Ejected jet particles are expected to induce the formation of a radio lobe that expands to give rise to a shock-compressed gas bubble populated by relativistic electrons, which should generate non-thermal emissions \citep{Scheuer1974, Begelman1989, Kaiser1997, Kaiser2004}. 
Large extended structures around powerful extragalactic radio
sources of type FR II \citep{Fanaroff1974} ascribed to this phenomenon have been observed for decades. Although, similar albeit smaller structures are expected around stellar-mass BHs \citep{Kaiser2004}, only a handful of examples exist. A possible explanation is that the formation of such features in the ISM may not be universally observed at each interaction site, this being subject to the properties of both the local environmental conditions (e.g. density and chemical composition) and the intrinsic properties of the jets (e.g. velocity, launch direction, duty cycle,  lifetime). Yet, the attributes of jet-ISM interaction regions encapsulate significant insights into poorly understood characteristics of jets, specifically the total jet power,  radiative efficiency, speed, and matter content.

Analogous to the studies of the radio lobes associated with AGN jets, the jet-ISM interaction regions can be used as calorimeters of the `power $\times$ lifetime' product of the jet.  This approach has already been adopted in the case of Cyg X–1. Deep radio observations of this source revealed a semi-circular shock region with a diameter of $\sim$5 pc aligned with the resolved AU-scale radio jet \citep{Stirling2001} that develops at the location where the collimated jet impacts the ISM, where a radio lobe is inflated \citep{Gallo2005}. To date, this structure and similar ones previously discovered near the peculiar system SS433 \citep{Begelman1980, Dubner1998} are the clearest examples of jet-inflated radio lobes near a BH X-ray binary (or microquasar) in the Galaxy (but see \citealt{Sell2015} for an alternative explanation for the case of Cyg X-1). A few more examples have also been found, including around the neutron star X-ray binary Cir X-1 \citep{Sell2010,Coriat2019}, in the vicinity of the peculiar BH X-ray binary in the nearby galaxy NGC 7793 \citep{Pakull2010, Soria2010, Hyde2017}, in the Large Magellanic Cloud \citep{Hyde2017}, and in NGC 300 \citep{Urquhart2019}.

\bigskip 

GRS~1915+105 is one of the best-studied Galactic BH X-ray binaries.  Located at a radio parallax distance of 9.4$\pm$0.7~kpc \citep{Reid2023}, GRS 1915+105 hosts a stellar-mass BH with mass $\sim$11.4\,\solarmass \citep{Reid2023}, which - when active - is believed to accrete erratically at a mass accretion rate close to its Eddington limit \citep{Done2004}. This system was the first Galactic BH observed to display relativistic, apparently super-luminal, radio ejections \citep{Mirabel1994}, and it is still regarded as the archetypal Galactic source of relativistic jets.

GRS 1915+105 appeared as a bright transient in August 1992 \citep{Castro-Tirado1992} and remained very bright in X-rays and radio until recently \citep{Motta2021}.
Over the past 30 years, it has alternated between extraordinary activity phases with strong and variable radio and X-ray emission, and long periods of reduced activity, with relatively low radio and X-ray fluxes, with the latter being associated with a hard state X-ray energy spectrum \citep{Belloni1997}. 
In such a hard state, the source is thought to be `jet-dominated’; that is, the majority of the liberated accretion power is in the form of a radiatively inefficient jet and is not dissipated as X-rays in the accretion flow \citep{Fender2003}. In radio, an extended milli-arcsecond scale core radio jet is commonly observed \citep{Dhawan2000}, which is variable on relatively short timescales ($<1$hr). Transitions to softer accretion states are usually accompanied by strong radio flares, which are associated with relativistic ejections that produce extended jets. Such jets have been repeatedly resolved, propagating out at approximately 30$^{\circ}$ from the north-south direction on the plane of the sky to distances between a few \citep{Dhawan2000, Rushton2007} to tens or hundreds of milli-arcseconds \citep{Fender1999, Miller-Jones2007}.

Searches for extended jet-ISM interaction structures around GRS 1915+105 at radio wavelengths have been unsuccessful \citep{Rodriguez1998}. Only two relatively compact regions with flat radio spectra symmetric about GRS 1915+105 - IRAS 19124+1106 and IRAS 19132+1035 - were identified as candidate interaction sites as early as the late 1990s \citep{Rodriguez1998}. Both of these sources are located approximately 17\,arcmin away from GRS 1915+105, at about 40\,pc if located at the same distance as GRS 1915+105, and their flat radio spectra are incompatible with synchrotron emission. The {H\sc{i}} distance to IRAS 19132+1035 places the hot spot at a distance of 6.6$\pm$1.4 kpc from the observer \citep{Chaty2001}, which is marginally consistent with the most recent parallax distance of GRS 1915+105 \citep[9.4$\pm$0.7~kpc,][]{Reid2023}.
\cite{Kaiser2004} developed a self-similar model for the large-scale jets from GRS 1915+105, applying a fluid dynamical model originally developed for extragalactic jets to this microquasar to infer the properties of the jet presumably responsible for the interaction \citep{Kaiser1997}. These authors concluded that while a bow shock structure should exist, it had not been observed around GRS 1915+105 (nor in the vicinity of other X-ray binaries) due to the limited sensitivity of the available observations. 
More recently, using ALMA line observations, \cite{Tetarenko2018} showed that the southern jet from GRS 1915+105 collides with a molecular cloud coincident with IRAS 19132+1035, which is likely heated by a young medium-mass star cluster. These results further supported the connection of IRAS 19132+1035 with the jets from GRS 1915+105, although they indicated that the interaction between the BHXB jet and the ISM in this region occurs on smaller scales than previously hypothesised. 

In this paper, we report on the discovery of a large-scale structure near GRS 1915+105 in sensitive MeerKAT observations with properties consistent with those expected if the jet developed a bow shock, as predicted by \cite{Kaiser2004}. 
The paper is structured as follows: Section~\ref{Sec:Obs} describes the observations we carried out and the data reduction. Section~\ref{Sec:results} describes the properties of the structure we identified in the images. Section~\ref{sec:analysis} reports the detailed calorimetry of the jets, which we employed to estimate the energy transported in the jets and transferred to the ISM. Section~\ref{sec:discussion} discusses our results and their implications. Finally, Sect.~\ref{sec: conclusions} contains our concluding remarks.

\section{Observations and data reduction}\label{Sec:Obs}

We observed GRS 1915+105 for a total of 65 times with the MeerKAT interferometer as part of the ThunderKAT Large Survey Project \citep{Fender2016b}, which between 2018 and 2023 routinely observed active X-ray binaries, cataclysmic variables, supernovae, and gamma-ray bursts.
We started observing GRS 1915+105 (J2000 19$^h$15$^m$11.56$^s$ +10$^{\circ}$56$'$44.9$''$) with MeerKAT soon after the start of operations, on 2018-12-08, and subsequently every several weeks until March 2020 \citep{Motta2021}. On March 18th, 2020 the AMI-LA interferometer, which has been carrying out a long-term monitoring of GRS 1915+105, interrupted operations due to the Covid-19 outbreak \citep{Motta2021}. Thus, we started weekly monitoring of the target with MeerKAT, which ended in April 2021. After this date, sparse MeerKAT observations of GRS 1915+105 were taken until June 2023. The observations used in this work are listed in Appendix \ref{sec:app_obs} and are all publicly available on the SARAO archive\footnote{\url{https://archive.sarao.ac.za/}}. 

We used the telescope’s L-band receivers and obtained data at a central frequency of 1.28 GHz across a 0.86 GHz bandwidth (856 – 1712 MHz). The correlator was configured to deliver either 4096 or 32768 channels, with an 8-second integration time per visibility point. In either case, data were binned down to 1024 channels for consistency before any further analysis. Between 58 and 64 of the 64 available dishes were used in the observations, with a maximum baseline of 7.698 km. 
The first MeerKAT observation (observing block 1544268663) consisted of a 90~min run, of which 60 min was on-source, 20 min on the primary (flux and bandpass) calibrator (J1939$-$6342), and 3 minutes on the nearby secondary (phase) calibrator (J2011$-$0644). 
All the other observations consisted of 15 minutes of on-source time, bookended by two 2-minute scans of the secondary calibrator and a 10-minute observation of a primary calibrator.

We conducted the subsequent analysis via a set of \textsc{Python} scripts specifically tailored for the semi-automatic processing of MeerKAT data (\textsc{OxKAT}\footnote{\url{https://github.com/IanHeywood/oxkat}}, \citealt{Heywood2020}). 
The \textsc{CASA} package \citep{CASAteam2022} was used to flag the first and final 100 channels from the observing band, autocorrelations and zero amplitude visibilities. 
Further flagging of the data was performed to remove RFI in the time and frequency domain. Flux density scale, bandpass and delay corrections were derived from the scans of the primary calibrator, while the complex gain and delay corrections were obtained from the scans of the secondary calibrator. A spectral model for the phase calibrator was derived starting from the flux and bandpass calibrator, by temporarily binning the data into 8 equal spectral windows.
The above corrections were applied within \textsc{CASA} to the target field, which was split to a different measurement set and further flagged using the \textsc{TRICOLOUR} package \footnote{\url{https://github.com/ska- sa/tricolour/}}, after averaging the data in time (8~s) and frequency (8 channels) for imaging purposes.

We used \textsc{WSClean} (\citealt{Offringa2012}) to obtain a first image of the entire square-degree MeerKAT field by combining a subset of 10 observations where GRS 1915+105 showed a relatively low flux density (below $\sim$10~mJy; see \citealt{Motta2021}). By combining several observations we obtained a better $uv$-coverage compared to that of individual observations, which is crucial for a good image deconvolution, while by selecting only observations where the target was relatively faint we mitigated the artefacts that may arise in imaging bright and/or variable sources with a sparse {\it uv}-coverage. We used the resulting image to generate a deconvolution mask, which we adopted to image each observation individually. Each measurement set was then self-calibrated using \textsc{CUBICAL} \citep{Kenyon2018} to solve for phase and delay corrections for every 32 seconds of data.

Since GRS 1915+105 showed dramatic flux density variations, spanning a dynamic range of over 2 orders of magnitude (i.e. from $\approx$1~mJy to over 600~mJy), we {\it uv}-subtracted it from each observation before combining several observations to avoid the effects of a strong and variable source near the phase centre\footnote{All observations have been pointed at a position slightly offset with respect to the nominal position of GRS 1915+105, to avoid artefacts that can sometimes arise at phase centre.} when building a deep image of the field. We first used \textsc{WSclean} to generate a model of the field with higher frequency resolution (16 channels instead of 8) for each observation. Then we partitioned the model into a second FITS cube containing only GRS 1915+105, which we subsequently subtracted from the model visibilities of the full sky. Finally, we used \textsc{WSclean} to combine all the observations, to obtain a deep image of the field without the point source corresponding to GRS 1915+105. We exclude from the final stacked data-set those measurement sets that individually yielded images affected by severe artefacts. 

\begin{figure*}
\centering
\includegraphics[width=0.99\textwidth]{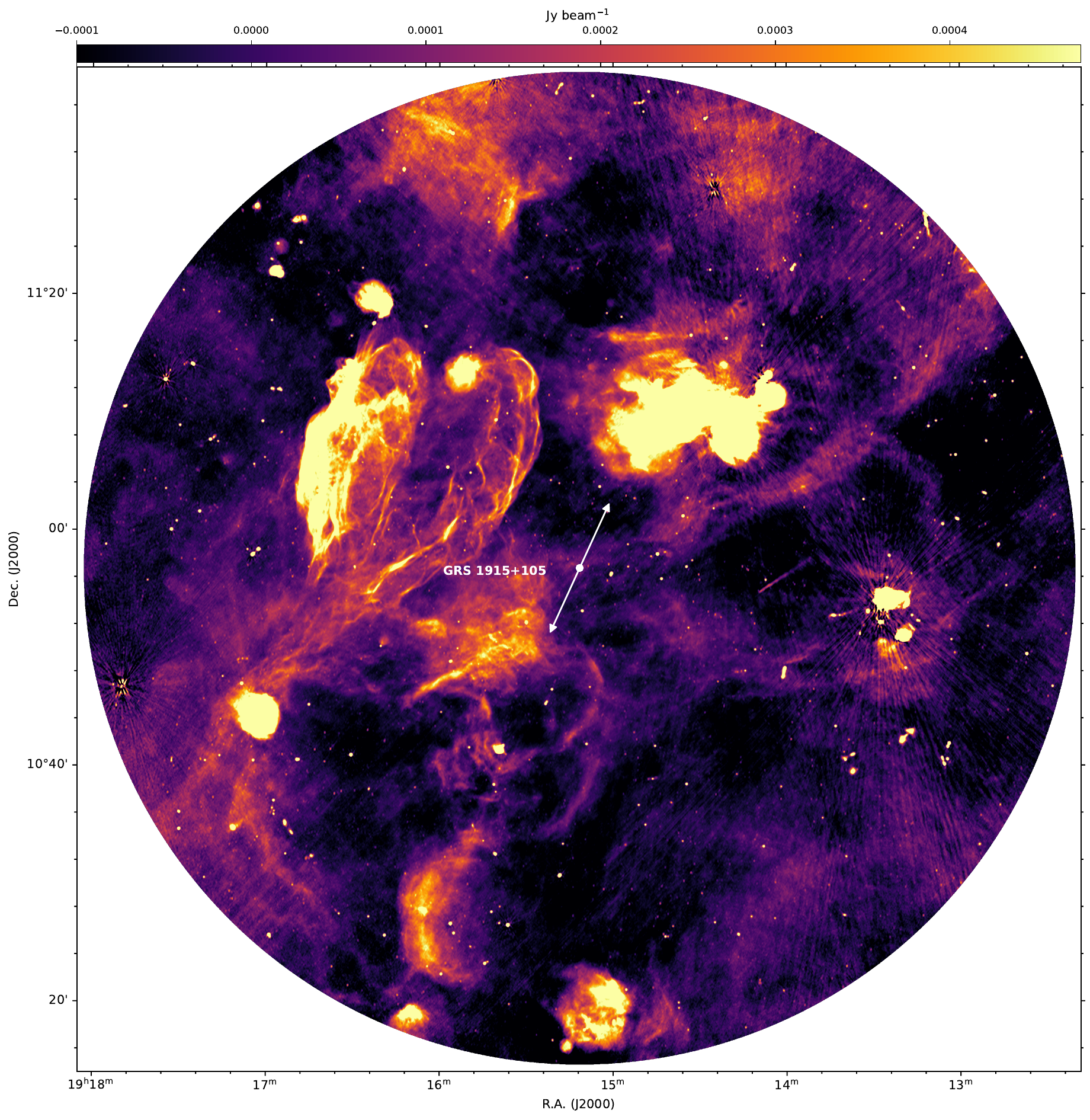}
\caption{Field of GRS 1915+105 as seen by the MeerKAT interferometer at 1.28 GHz in a total on-source time of 15.5 hr. The restored beam for this map is circular and has a radius of 6.6". The circle marks the position of GRS 1915+105, and the two arrows mark the direction of the jets identified in the '90s \citep{Fender1999}. In the image, north is up and east is left. }
\label{fig:1915_field}
\end{figure*}

The resulting map has been primary beam corrected using the \textsc{katbeam} package\footnote{https://github.com/ska-sa/katbeam}, blanking the map beyond the nominal 30\% level\footnote{Corrections are applied to areas within the primary beam where the response remains above 30\% of its peak value. Where the beam response falls below 30\% of its maximum the signal-to-noise ratio drops, making it challenging to distinguish between actual signals and noise. Hence, applying corrections in these regions could amplify noise more than the signal, leading to unreliable data.}.
The resulting wide-field image is presented in Fig. \ref{fig:1915_field}, which constitutes the deepest radio image of this field in L-band at the time of writing, with an overall exposure of 14 hours. 

\begin{figure}
\centering
\includegraphics[width=0.48\textwidth]{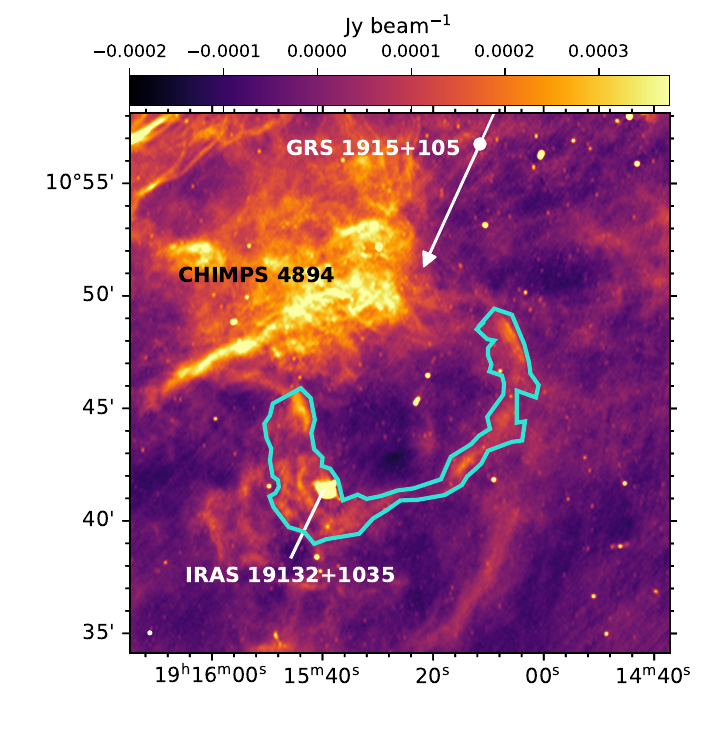}
\caption{Zoom-in of the jet-ISM interaction region. The bow shock structure is marked by the cyan region, which has been defined by eye. The position of GRS 1915+105 and IRAS 19132+1035, as well as the jet direction, are marked.}
\label{fig:1915_zoom}
\end{figure}

\section{Results}\label{Sec:results}


The MeerKAT image of the GRS 1915+105 field is presented in Fig. 1. The field is complicated with multiple large-scale emission regions and compact sources. GRS 1915+105 is in the middle of the field, represented by the white dot. The direction of the radio jets, which have been repeatedly resolved as extended radio jets on a range of scales from $\sim$1 milli-arcsecond to hundreds of arcseconds \citep{Dhawan2000,Miller-Jones2005,Rushton2007,Fender1999,Miller-Jones2007}, is denoted in Fig. 1 with white arrows. We discover an arched structure that appears approximately 17 arcmin to the south-east of the GRS 1915+105 position, with an apparent diameter of  10~arcmin, and an average flux density between 0.1 and 0.2 mJy (see Fig. 2). The arch appears to be connected with a large H{\sc ii} region (CHIMPS 4894\footnote{\url{http://simbad.cds.unistra.fr/simbad/sim-id?Ident=CHIMPS+4894\&}.}; see Fig. \ref{fig:1915_zoom}) that extends on either side of the bright extended source IRAS 19132+1035 and aligns with the direction of the extended radio jets from GRS 1915+105. These properties suggest that the structure we observe is compressed material behind a bow shock, which may have formed at the head of a large-scale dark jet (i.e. not directly imaged in the radio band) that rams into dense ISM material, carving out a cavity in the ISM. 

The observed configuration is reminiscent of what was observed around Cyg X-1, where a bow shock formed as a large-scale jet connected with the tail of a nearby HII nebula \citep[Sh2-101, or the Tulip nebula, ][]{Gallo2006}. The large-scale dark jet in this case was able to blow a cavity into the ISM with a diameter of approximately 5 pc, at a distance of 6 pc from the BH\footnote{Assuming a distance of $\sim$ 2.22~kpc, \cite{Miller-Jones2021}.}. For the case of Cyg X-1, \cite{Sell2015} stressed that the possibility of a massive O-star wind playing a role in inflating the bubble marked by the bow shock cannot be ruled out. In principle, a similar situation could exist for GRS 1915+105, for which we have no final proof that the structure we observe is really connected with a bow shock due to the jets, and not a fore/background structure. However, we note that the companion star in GRS 1915+105 is not massive, but an evolved low-mass companion \citep{Reid2014}, which is not likely to generate strong winds. 

Based on the above, in this work, we postulate that the structure we discover in the MeerKAT data is induced by the jet interacting with the surrounding medium. Under this assumption, we note that the cavity blown by the jet into the ISM 
is significantly larger than in the case of Cyg X-1. At a distance of $\approx$9.4 kpc and assuming a source inclination angle $i$ of 60$^{\circ}$ \citep{Reid2014}, the bubble near GRS 1915+105 has a physical diameter of approximately 30 pc and formed $\approx$42 pc away from the binary, and the associated arched structure is 16 times more luminous than that near Cyg X-1. 

\bigskip 

The postulated jet-ISM interaction site near GRS 1915+105 shows a complex structure formed by various areas. 
In Fig. \ref{fig:schem} (panel A), we show a sketch of the region marking its various components, which we also describe below:
\begin{itemize}
    \item IRAS 19132+1035 (henceforth \textit{the IRAS region}), which emits flat-spectrum thermal radiation \citep[see][]{Rodriguez1998}. This feature is also characterised by molecular line emission, which indicates that the gas in the region is a mix of hot and cold gas, the latter possibly hosting young massive stars, which could contribute significantly to the emission from the region \citep{Tetarenko2018}. 
    \item A non-thermal feature (henceforth \textit{the northern feature}) to the north-east edge of IRAS 19132+1035 pointing towards GRS 1915+105. This region is the site of particle acceleration that originates at the end of the large-scale jet. This structure was already tentatively associated with the jet from GRS 1915+105 in the late '90s \citep{Rodriguez1998} and \cite{Tetarenko2018} confirmed the association by showing the presence of molecular lines near this region, consistent with the effect of an active jet pushing into cold gas. The northern feature emits steep-spectrum synchrotron radiation \citep[see][]{Rodriguez1998}, and in a AGN would be identified with a hot-spot \citep{Kaiser2004}.
    \item An arched structure (henceforth \textit{the bow shock structure}) with an average brightness of 0.15 mJy/beam that extends to either side of the IRAS region. This feature, which was not known before and has been discovered by MeerKAT, likely marks the presence of shock-compressed material and of a bow shock. In the old VLA data \citep[][]{Rodriguez1998} the structure was not detected due to limited sensitivity, although \cite{Kaiser2004} predicted it, and estimated that the brightness of the feature should have been of the order 0.1 mJy/beam. This brightness level is well below the sensitivity of the VLA observations, but remarkably consistent with what we observed with MeerKAT. The superior surface brightness sensitivity of MeerKAT is also likely responsible for the detection of the dim diffuse emission that other instruments would resolve out.
\end{itemize}

\section{Analysis}\label{sec:analysis}

To interpret the interactions of the large-scale jet of GRS 1915+105 with its environment, we employed the model originally developed by \cite{Kaiser1997} for the jets of radio galaxies and later adapted to Galactic sources by \cite{Kaiser2004}. We assumed that the supersonic jets emerging from a central source end in strong shock fronts at the location where they impact the ambient medium. The plasma from the jets will inflate two lobes that are overpressured with respect to the environment, and the pressure in the lobes should be high enough that the jets are confined while propagating through this region so that the energy transported by the jet is deposited at the jet end shock without significant losses. The lobes, therefore, expand forwards and sideways, and thus a bow shock forms, while the jets -- expanding in a low-density but high-pressure medium -- are protected against disruption due to the turbulence in the gas in the environment. The jet direction is assumed to remain constant over time, as the model requires that the jets propagate in an environment that has been cleared out by previous ejections.

\begin{figure}
\centering
\includegraphics[width=0.45\textwidth]{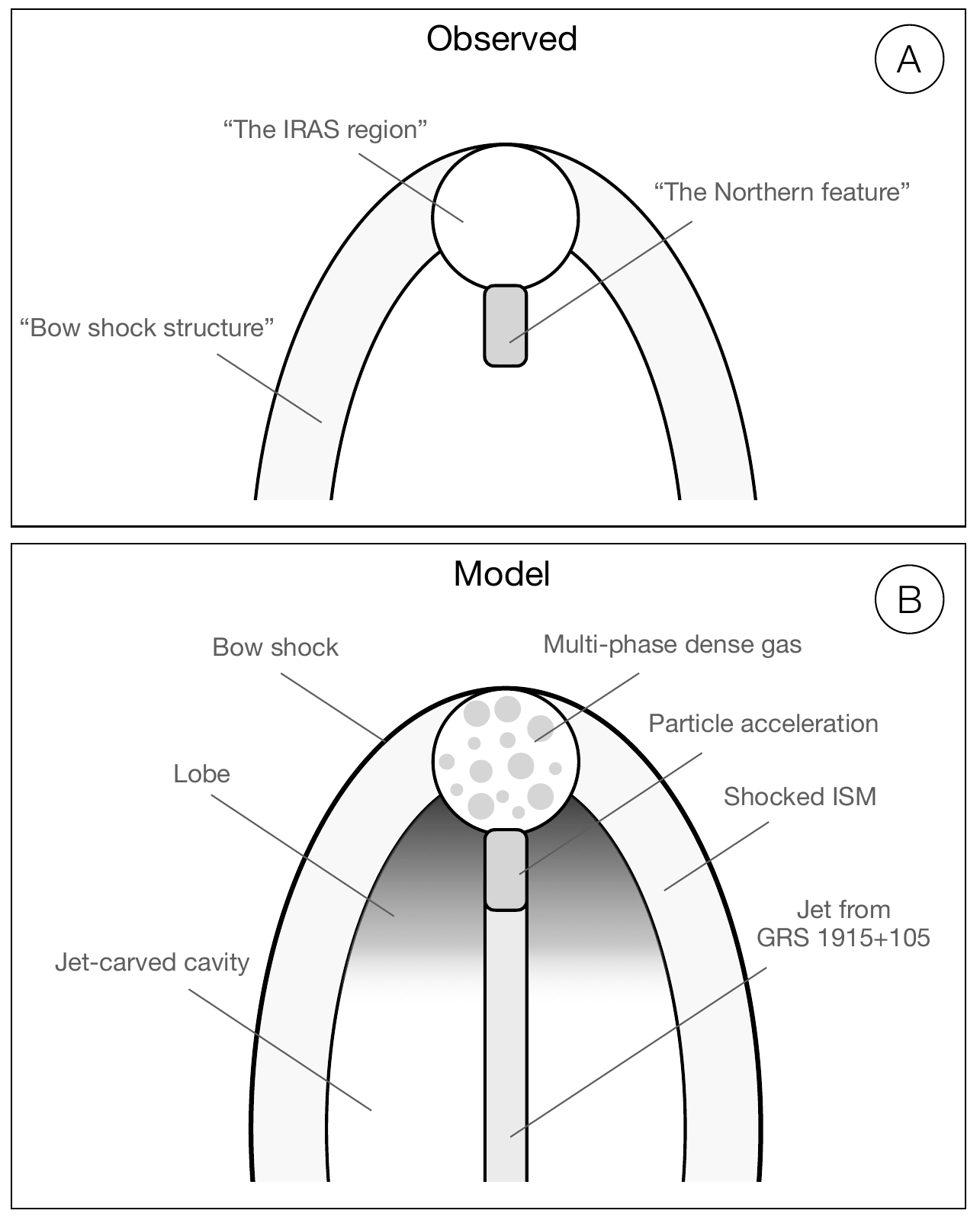}
\caption{\textit{Panel A:} Schematic picturing the jet-ISM interaction region as observed in the MeerKAT image. \textit{Panel B:} Sketch of the model employed to interpret the observed structure.}
\label{fig:schem}
\end{figure}

The sketch in Fig. \ref{fig:schem} (panel B) shows the model we employ to interpret the structure we observed in the MeerKAT image. The site of particle acceleration at the end of the jet, the excited multi-phase gas which the jet interacts with, and the shocked compressed material are visible in the radio band as the northern feature, the IRAS region, and the bow shock structure, respectively. The large-scale jet from GRS 1915+105 and the lobe fed by the jet are not visible, possibly due to sensitivity reasons. This morphology could be seen as a small-scale analogue of what is seen around, for example, Cygnus A \citep{Carilli1991}. 

With this model in mind, we could determine the properties of the jet. We proceeded as follows:
\begin{enumerate}
    \item We first derived the density of the shocked material at the end of the jet by assuming that the jet hits an overdense region, corresponding to the IRAS region, and by assuming that the continuum emission is predominantly Bremsstrahlung. Then, we derived the pre-shock ISM density. 
    \item We modelled the non-thermal northern feature assuming that the continuum emission is predominantly due to synchrotron and derived the minimum energy magnetic field and pressure.
    \item We modelled the bow shock structure assuming that the emission is due to synchrotron in analogy to what is done for the northern feature, and we derived the minimum energy magnetic field and pressure within the lobes. For completeness, we also considered the case where the emission is due to Bremsstrahlung. 
    \item We derived the jet characteristic scale to verify the applicability of the jet model by \cite{Kaiser2004}.
    \item We estimated the jet advancing velocity assuming that it equals the velocity of the bow shock in the ISM and derived the jet age. 
    \item We estimated the jet energy transport rate.
    \item Finally, we derived the jet energy transfer rate via two more simplified calorimetry methods based on the estimate of the enthalpy of the jet-ISM system and on a different use of the minimum energy argument previously applied to AGNs. 
\end{enumerate}

For all the quantities we derive below, we provide either ranges or lower limits, depending on the assumptions made. Ranges were obtained using a Markov chain Monte Carlo approach in order to derive a posterior distribution for each given quantity, from which we obtained an upper and lower boundary as the values corresponding to the 68\% confident range (16th and 84th percentile) unless otherwise specified. This choice was dictated by the large uncertainties on some of the parameters we used or by the need to make conservative assumptions. 
The calculations behind the treatment below are presented in the accompanying Jupyter notebook. 

\subsection{Assumptions}\label{sec:assumptions}

In the interest of clarity, in this section we state the assumptions that are made in our analysis.
This work hinges on two major assumptions, which have been already introduced above. The first is that the IRAS region emits radiation at least partly due to interaction with a jet from GRS 1915+105. The second assumption is that the arched structure we discovered in the MeerKAT images is the signature of a bow shock induced by the jet from GRS 1915+105, marking the edge of a lobe. 
Other assumptions we made are listed below and discussed in the following sections: 
\begin{itemize}
    \item The IRAS region is filled by multi-phase gas, i.e. a mix of hot, fully ionised pure hydrogen gas at a temperature between 10$^4$\,K and 3$\times$10$^6$\,K, and cold, denser molecular gas with temperatures around $10-100~{\rm K}$.
    \item The emission from the hot gas in the IRAS region is predominantly Bremsstrahlung, as indicated by the flat continuum spectrum, \citep[][]{Rodriguez1998}. 
    \item The IRAS region can be modelled as a sphere with a certain filling factor to account for the inhomogeneity of the gas (see Sec \ref{sec:iras}). 
    \item The emission from the northern feature is predominantly synchrotron radiation, as indicated by the steep continuum spectrum from this region, \citep[][]{Rodriguez1998},  with a spectral index $\alpha$= -0.75. 
    \item The northern feature can be modelled as a cylindrical shape.
    \item The gas within the northern region is in equipartition (see Sect. \ref{sec:cylinder}).
    \item The lobe blown by the jet, partly delimited by the bow shock structure, can be modelled as an ellipsoidal shape with a major axis equal to the distance between GRS 1915+105 and the IRAS region and minor axes equal to the bow shock width (see Sect. \ref{sec:lobes}).
    \item The jet opening angle is constrained between 0 and 10 degrees.
    \item The adiabatic indices of the material in the lobe cavity, in the external medium, and within the jet ($\Gamma_j$) at the jet-ISM interaction site are assumed to equal 5/3 (see Sect. \ref{sec:calorimetry}).
\end{itemize}

\subsection{The structure of the jet--interstellar medium interaction site}\label{sec:Largejet}

    \subsubsection{The IRAS region}\label{sec:iras}

The continuum emission of IRAS 19132+1035 is characterised by a flat radio spectrum, which indicates a Bremsstrahlung origin from a gas emitting at a temperature of T $\sim 10^4K$. Below this temperature, hydrogen is not significantly ionised and hence cannot produce Bremsstrahlung radiation efficiently. This is further supported by the detection of a H92$\alpha$ recombination line with width $\sim$25 km\,s$^{-1}$ \citep{Rodriguez1998}, which implies a temperature of at least T $\sim$10$^4K$. We assumed a conservative upper limit of T = 3$\times$10$^6~K$ to the temperature of the Bremsstrahlung emitting gas, assuming that the shock forming at the jet end is radiative (and thus the electrons need to have sufficient energy to emit photons effectively when decelerated) and noting that Bremsstrahlung cooling becomes significant at temperatures typically above $10^6 K$ for particle densities typical for the ISM and HII regions (0.1-100 cm$^{-3}$; see e.g. \citealt{Gould1980}).

From the monochromatic emissivity $\epsilon_\nu$ of ionised hydrogen due to thermal Bremsstrahlung \citep{Longair1994}, we can derive the electron density $n_e$: 

\begin{equation}\label{eq:electronD}
n_e = \sqrt{ \frac{\epsilon_\nu}{g(\nu, T) \sqrt{T} C_{radio} \exp{(\frac{h\nu}{k_b T})} }},
\end{equation}
where $ C_{radio} = 6.8\times 10^{-38}\, {\rm erg\,s}^{-1}\,{\rm cm}^{-3}\,{\rm Hz}^{-1} $, and $g(\nu)$ is the Gaunt factor (see Appendix \ref{sec:App_IRAS} for a detailed treatment). We assume that the gas in the region is pure hydrogen, and thus the number density of electrons $n_e$ equals the number density of protons $n$, and that the density and temperature are uniform within the volume $V$ considered, which in the case of the IRAS region for simplicity we modelled as a sphere with apparent diameter 36 arcsec. Since the IRAS region contains also cold and dense molecular gas, which does not contribute to the continuum emission, we assume a sphere filling factor of 0.5 to account for the gas inhomogeneity. 

Considering a temperature range of T $\sim 10^4 - 3\times 10^6~K$, we derived an electron density of the shock-compressed gas between 395 and 630 cm$^{-3}$. At the above temperatures, such particle densities imply that hydrogen must be fully ionised, and thus the electron density equals the proton density. We note that our approach differs from that taken in \cite{Gallo2005} for the case of Cyg X-1, who assumed an ionisation fraction of $\approx$2\%, which resulted in a significantly higher total particle density \citep[see also][submitted]{Atri2025}. 
Assuming that the shock-compressed gas is approximately 4 times denser than the surrounding medium, we deduce that the pre-shock gas density (equal to the un-shocked ISM electron density) is between $\approx$100 and 160 particles per cm$^3$. This range is consistent with the typical density of an HII region (and over 10 times larger than the typical ISM density). Since the bow shock structure seems to be indeed connecting with an HII region, these density values are reasonable.

    \subsubsection{The northern feature}\label{sec:cylinder}

Both the MeerKAT radio map and the older VLA radio maps \citep[see][]{Rodriguez1998} of IRAS 19132+1035 show an elongated feature extending from the main structure towards the direction of GRS 1915+105 for approximately 17.4 arcsec, with a steep radio spectrum and an integrated flux density of 1.57 mJy. In the VLA images the spectrum of the feature is consistent with a power law F $\propto$ $\nu^\alpha$ with $\alpha$ = -0.75, typical of synchrotron emission, as opposed to the flat spectrum of the IRAS region. The synchrotron spectrum implies the presence of a relativistic plasma consisting of a magnetic field and a population of relativistic electrons, for which we can calculate the total energy, and from there infer the magnetic field corresponding to minimum energy, and the associated minimum energy pressure.

We assumed a radio spectrum with some spectral index $\alpha$ between upper and lower frequencies $\nu_1, \nu_2$, and $\alpha = (1-p)/2$ = -0.75 \citep{Mirabel1994}. 
The energy density in magnetic fields is $B^2 / 8\pi$ (in cgs units); therefore, for a volume $V$, the total energy in the magnetic field is
\begin{equation}\label{eq:EB1}
E_B = \frac{B^2}{8 \pi} f V,
\end{equation}
where $f$ is a filling factor, which represents the fraction of the physical source's volume that is occupied by the magnetic field and relativistic particles. In our case, we assume $f$ = 1.

Given a population of electrons with a power law distribution with index $-p$ between energy limits $E_1$ and $E_2$, the total energy in these electrons can be written as
\begin{equation}\label{eq:Ee}
E_e =  c_{12}(p,\nu_1,\nu_2) B^{-3/2} L,
\end{equation}
where we characterised the synchrotron emission as coming from a single frequency $\nu = c_{12} B E^2$ for a given electron energy and magnetic field. The pseudo-constant $c_{12}$ encapsulates all the information about the frequency range and the slope of the energy spectrum between those limits. In Appendix \ref{sec:App_IRAS} a more detailed treatment is provided, which includes the explicit form of the $c_{12}$ constant. 

Since the energy stored in the magnetic field is proportional to the magnetic field strength squared, there must be a minimum of the total energy as a function of $B$, if we assume that there are no other contributions to the energy budget. Such a minimum occurs close to the equipartition of the energy,
%
which corresponds to an equipartition magnetic field $B_{eq}$ given by
\begin{equation}\label{eq:Beq}
    B_{\rm eq} = \left(6 \pi \frac{\eta}{f} c_{12} \frac{L}{V}\right)^{2/7},
\end{equation}
where $\eta$ indicates how much energy is stored in the protons that accompany the electrons (in the case of a normal baryonic plasma), and a total pressure is given by
\begin{equation}\label{eq:peq}
    p_{\rm eq} = \frac{7}{9} \frac{B^2_{\rm min}}{8\pi}(k+1) ,
\end{equation}
with $k$ equal to the ratio of the internal energy stored in any particles that do not contribute to the synchrotron emission to the sum of the energy in the magnetic field and in the relativistic electrons.

For simplicity, we assume a cylindrical geometry for the northern feature, with a radius of 3.9 arcsec and a length of 17.4 $\times$ sin($i$)$^{-1}$ arcsec \citep{Mirabel1994}, where $i$ is the angle of the long axis of the cylinder to our line of sight (which we assumed to be  60$^{\circ}$). Thus, the volume of the northern feature is V$_{cyl}$ = 2.6$\times 10^{54}$ cm$^3$, and its luminosity $L_{cyl}$ =  2$\times 10^{29}$erg\,s$^{-1}$. The predicted minimum energy magnetic field is B$_{min}$>  2.2$\times$10$^{-5}$G, and the minimum energy pressure is $p_{B min}$ = 1.5$\times$10$^{-11}$ erg/cm$^{3}$. 

    \subsubsection{The lobe}\label{sec:lobes}

In the case of jets from AGNs, a substantial fraction of the energy transported by the jet is transferred to a population of relativistic electrons and a magnetic field in the lobes, which become bright due to synchrotron emission. The lobes are thus generally easily detected in radio in the case of AGNs, and similar lobes may be expected to emerge around microquasars due to a similar mechanism if the environment density is large enough \citep[see also][]{Heinz2002}. The search for such lobes around GRS 1915+105 has not been successful so far, likely due to the insufficient sensitivity of such searches. In this regard, \cite{Kaiser2004} argued that, under a reasonable set of assumptions, a synchrotron lobe would have had a specific intensity of approximately 0.1 mJy/beam at a resolution of 4 arcsec (i.e. comparable with the MeerKAT resolution), which was below the 1-$\sigma$ rms noise of the VLA data taken in the '90s. Such a brightness is remarkably consistent with that we measured in the bow shock structure in the MeerKAT data, i.e. 0.15 mJy/beam. 

Let us assume that the bow shock structure corresponds to the brightened limb of the lobe, and let us estimate the magnetic field implied by the measured average flux density and the associated pressure inside the lobe, which we can compare with the pressure within adjacent jet regions. 
We assume that the conditions of minimum energy hold in the lobe, and so we can infer the amount of energy necessary to generate the detected luminosity, and from there the magnetic field corresponding to the minimum energy criterion B$_{min}$ using Eq. \ref{eq:Beq} \citep[see also e.g.][]{Longair1994}. We neglected any energy losses of the relativistic electrons due to radiation processes. The minimum energy magnetic field B$_{min}$ can be compared with the magnetic field density we obtained for the non-thermal northern feature (see Sect. \ref{sec:cylinder}). Due to the low luminosity of the bow shock we cannot directly measure its spectral slope $\alpha$. Hence, following \cite{Kaiser2004}, based on the fact that the relativistic plasma powering the northern region inflates the cavity, we assume that also the bow shock structure emits synchrotron radiation with a spectral slope $\alpha = -0.75$. 

We model the radio lobe as an ellipsoidal shape, with a major axis equal to the distance between GRS 1915+105 and the outermost edge of the IRAS region, and a semi-minor axis equal to the width of the bow shock in the direction orthogonal to the jet (see Table \ref{tab:BHpar}). We obtain a magnetic field density of B$_{min}^{lobe}$$\sim$6.4$\times$10$^{-6}$\,G, and a minimum energy lobe pressure is p$_{lobe}$ $>$ 1.3$\times$10$^{-12}$erg cm$^{-3}$, which are consistent with the values we obtained for the northern feature. The pressure is also significantly larger than the pressure in the unperturbed ISM, as expected based on our assumptions.

If, instead, we assume that the bow shock structure emits Bremsstrahlung radiation similarly to the IRAS region, we can infer the implied electron density using Eq. \ref{eq:electronD} following the procedure described in Sect. \ref{sec:iras}. In this case, we assume that the depth of the ring is of the order of the projected width of the bow shock on the plane of the sky ($\Delta R$) because of limb brightening, and we determine the unit volume associated with the average specific intensity of the region (0.15~mJy/beam) as the beam surface area (6.6 arcsec$^2$) times the ring thickness (60 arcsec, i.e. 2.7 pc). The associated pre-shock electron density ranges between 13 and 26 particles per cm$^3$, i.e. approximately a factor of 6 lower than what we inferred in Sect. \ref{sec:iras}.

\subsection{Jet calorimetry}


The jet model assumes a constant energy transport rate, $Q_0$, for each side of the jet, and is averaged over the lifetime of the jets. We further assume that the ambient medium has a constant density $\rho_0$. Under these circumstances the evolution of the large-scale structure created by the jet (lobe and bow shock), is self-similar once the jet extends beyond its characteristic length scale, which is given by Eq. 5 in \cite{Kaiser2004} \cite[see also][]{Falle1991}. 


The length scale depends on the density of the gas through which the jet is propagating, and the jet Lorentz factor $\gamma_j$. We derived the former in Sect. \ref{sec:iras}. The latter is more difficult to estimate \cite[see e.g.][]{Fender2003}, as many assumptions that can scarcely be verified are generally involved. For instance, the compact and steady jets (which are rarely resolved) may be fundamentally different from the discrete relativistic ejections more often observed around X-ray binaries, and hence the motion of radio ‘knots’ in the jets may simply reflect the motion of shocks along the jets rather than the bulk motion of the jet material. Any estimate of $\gamma_j$ will also depend on the distance to the source and on the orientation of the jet to our line of sight, which, when known, is generally affected by large uncertainties. In the interests of generality, here we use a conservative lower limit $\gamma_j$ = 1.02 (corresponding to $\beta$ = 0.2), which maximises L$_0$.

For a reasonable value of Q$_0$ $\sim$10$^{36}$ erg\,s$^{-1}$ (see below) and for the minimum density derived in Sect. \ref{sec:iras}, L$_0$ is of the order  10$^{-5}$ pc, that is,  orders of magnitude smaller than the jet length L$_j$ (> 40 pc), which confirms the applicability of the model by \cite{Kaiser1997}, as already showed by \cite{Kaiser2004}. 

    \subsubsection{Jet velocity and age}\label{sec:jetage} 

\cite{Kaiser1997} showed that the velocity of the jet advancing into the ISM is roughly equal to the expansion velocity of the bow shock driven by the jet itself, which implies that the age of the bow shock equals the age of the jet that produced the (most recent) shock. For a strong shock in a monatomic gas, the expansion velocity $\dot{L}$ is set by the temperature of the shocked gas, and is given by 
\begin{equation}\label{eg:Ldot}
 \dot{L} = \sqrt{\frac{16 k_b}{3 m_p}T},
\end{equation}
where $m_p$ is the mass of the proton. If the shock emits radiation, the initial post-shock temperature can be higher than that of the thermalised, bremsstrahlung-emitting gas. We assume that the temperature of the gas is constrained between 10$^4$ and 10$^6$ K, based on the similarity with the bow shock structure observed near Cyg X-1 (see \citealt{Gallo2005}, and accompanying paper by \citealt[][submitted]{Atri2025}). We estimate a shock velocity between 20 and 360 km\,s$^{-1}$, which implies that the bow shock will be highly supersonic with respect to the unshocked ISM. 

For an ISM with constant density \cite{Kaiser2004} showed that the growth of the jet length within the lobe as a function of time $t$ is given by
\begin{equation}\label{eq:jetL}
    L_j = C_1 ~\left(\frac{Q_0}{\rho_0}\right)^{1/5}~ t^{3/5},
\end{equation}
where $\rho_0$ is the density of the un-shocked surrounding medium, and $Q_0$ is the time-averaged jet power being transferred to the ISM. $\rho_0$ can be inferred from  Eq. \ref{eq:electronD}. $C_1$ is a dimensionless constant between 2 and 9 and median $\approx 3$ that depends on the thermodynamical properties of the jet material, and on the aspect ratio, R, of the lobe inflated by the jet. The explicit form of $C_1$ is given in Appendix \ref{sec:C1}. 
Taking the time derivative of Eq. \ref{eq:jetL}, one obtains
\begin{equation}\label{eg:jetAge}
    \dot{L}_j = \frac{3 L_j}{5 t},
\end{equation}
which implies
\begin{equation}\label{eq:jetT}
    t = \frac{3 L_j}{5 \dot{L_j}}.
\end{equation}

From Eq. \ref{eg:jetAge} we infer the jet age by assuming that $L_j$ corresponds to the distance between GRS 1915+105 and the outermost edge of the IRAS region, corrected by the jet inclination angle $i$ with respect to the line of sight. Based on our assumptions, the jet age is between 0.09 and 0.2 Myr, with an absolute upper limit of 1.3 Myr. This value essentially only depends on the shock advancing velocity, which in turn is calculated from the gas temperature $T$. We note that these estimates are robust as they are set by conservative assumptions, although better constraints could be obtained by refining the temperature range considered in the calculations. 

    \subsubsection{Jet power}\label{sec:calorimetry}

Assuming that the direction of the jet remains constant over time,  and again that the jet is colliding with a medium of density $\rho_0$, the power Q$_{jet}$ being transported by the jets into the surrounding averaged over its lifetime depends on the properties of the jet itself. Combining Eqs. \ref{eq:jetL} and \ref{eg:jetAge}, we obtained the jet power: 

\begin{equation}\label{eq:Qjet}
   Q_{jet} = \left(\frac{5}{3}\right)^3~\frac{\rho_0}{\bar{C_1}^5}L_j^2 ~\dot{L_j}^3.
\end{equation}

This expression depends strongly on the constant $\bar{C_1}$, which encapsulates several properties of the jet, including the jet opening angle $\theta$, which is the main driving parameter, the aspect ratio $R$ of the lobe inflated by the jet, and the adiabatic indices of the jet, lobe cavity, and un-shocked ISM (see \ref{App:App1}, \ref{sec:C1}). The indices can be assumed to be all equal to 5/3 \citep[see][for a discussion]{Kaiser1997,Kaiser2004}. Since changes in the indices only mildly affect the value of the constant, we choose to ignore possible variations.

For extragalactic jet sources, $R$ is in general lower than 2 \citep[][]{Kaiser2004}. In the case of the lobe in GRS 1915+105, assuming an ellipsoidal jet lobe, the aspect ratio $R$ of the lobe equals the axial ratio (given by the length of the lobe divided by its width), which gives $R$ = $R_{ax} \approx$ 49~pc/25~pc $\sim$ 2 (see also Appendix \ref{App:App1} and Sect. \ref{sec:App_northern}). 
This in principle provides a constraint on the jet opening angle since the variable $\bar{C_1}$ is a function of the opening angle (see Appendix \ref{App:App1}). In particular, $\bar{C1}$ $\approx$ 2 implies a jet opening angle of approximately 30 degrees. This value is large and at odds with the opening angles typically inferred for the transient jets seen in X-ray binaries, which are estimated or measured to be at most 4 degrees \citep[see e.g. ][]{Miller-Jones2006}. However, it is unclear whether the large-scale jet interacting with the ISM at large distances from the launching site has the same properties as either the transient intermediate-state jets often observed as a resolved radio-bright outflow expanding outwards on milli-arcsecond to arcsecond scales from the binary position or the continuous hard-state jets resolved at a few milli-arcsecond scales \citep{Fender2009}. Thus, for the sake of generality and to avoid extreme scenarios (i.e. very high or low collimation), we assumed a jet opening angle ranging between 1 and 10 degrees\footnote{The lower limit is set based on the assumption that a very low opening angle would require a mechanism to prevent the jet expanding sideways.}. We caution the reader that the choice of opening angle heavily affects the final jet energy estimate as larger opening angles correspond to increasingly more powerful jets, and thus it constitutes an important caveat in our treatment. 

For an opening angle between 1 and 10 degrees, $\bar{C1}$ ranges between 3.5 and 7.5, where the latter value comes from smaller opening angles. The one-sided (only one side of the bi-polar jet is considered) \textit{time-averaged} jet energy transferred to the environment Q$_{jet}$ from Eq. \ref{eq:Qjet} is Q$_{jet}$ $\sim$  3.3$\times$10$^{37}$ - 1.5$\times$10$^{39}$ erg\,s$^{-1}$, where the upper limit is largely driven by the upper limit imposed to the jet opening angle.

The pressure inside the lobe is given by

\begin{equation}\label{eq:pjet}
 p = 0.0675 \frac{C_1^{10/3}}{R^2}\left(\frac{\rho_0 Q_0^2}{L_j^4}\right)^{1/3},
\end{equation}
which ranges between 2.8$\times$10$^{-11}$ and 7.8$\times$10$^{-10}$ erg cm$^{-3}$. This means that the lobe pressure is consistent with the pressure we obtain for the cylindrical feature stemming from IRAS 19132+1035 (see Eq. \ref{eq:peq}), and significantly overpressured with respect to the un-shocked ISM, as required by the jet model. For a temperature of 50K, a typical ISM pressure would be of the order 9$\times$10$^{-13}$ erg\,cm$^{-3}$ assuming ideal gas conditions.



\subsubsection{The \textquoteleft Enthalpy\textquoteright\ method}\label{sec:enthalpy}

The estimate of the time-averaged energy rate transported by the jet described above hinges on the assumption that the emission from the IRAS region is Bremsstrahlung. In order to confirm our results we derived the energy deposited into the ISM via two more simplified though very generic calorimetry methods. 

In the case of AGNs, the jet power transport rate is often estimated by calculating the enthalpy of the regions powered by the jets, and dividing by an appropriate timescale \citep[e.g.][]{birzan2008,cavagnolo2010}.
A change in enthalpy at constant pressure is given by $dH = dU + pdV$, and it includes the internal energy stored in a volume and the work done in excavating it. Hence, it measures the total energy/heat ratio input by the jet. 
The enthalpy of an ideal gas with adiabatic index $\gamma$ is given by 
\begin{equation}
H = U + P V = \frac{\gamma PV}{(\gamma -1)},
\end{equation}
which for a relativistic ideal gas (i.e. $\gamma=4/3$), as is commonly assumed in the AGN case, gives $H= 4 PV$ (for a non-relativistic gas, $\gamma=5/3$ and $H=5PV/2$). The jet power is thus given by 
\begin{equation}
Q = \frac{4PV}{t},
\end{equation}
where $P$ is the pressure, which we assumed is equal to the minimum energy condition pressure we calculated above for the lobes (see Sect. \ref{sec:lobes}), namely, $1.3\times 10^{-12}~{\rm erg~cm^{-3}}$. The term $V$ is the volume excavated by the jet from GRS 1915+105, and $t$ is the jet age. 
We assumed that the cavity excavated by the jet is an ellipsoidal shape with a length equal to the source-to-bow shock distance, and a width equal to the bow shock diameter, which has a volume of 6$\times 10^{59}$cm$^3$. 
The upper limit to the jet age $t$ is around $0.4$ Myr (95th percentile) from the self-similar estimate (see Sect. \ref{sec:jetage}), which is consistent with the time scale obtained just by dividing the jet length by the shock velocity. The resulting one-sided jet transferred power ranges between $1.2\times 10^{37}~{\rm erg~s}^{-1}$ and $1.5\times 10^{39}~{\rm erg~s}^{-1}$. Despite the large uncertainty on this estimate, the value we obtained (especially its lower limit) is consistent with our estimate from the self-similar model in Sect. \ref{sec:calorimetry}. We note that these values could easily increase due to the presence of a non-radiating pressure component (and/or a non-volume-filling synchrotron plasma) in the lobes.

\subsubsection{The \textquoteleft Hot-spot \textquoteright \ method}\label{sec:hotspot}

For the case of AGNs, under the assumption of equipartition, \cite{Godfrey2013} derived  the jet power being supplied to the hot spot, which is given by
\begin{equation}\label{eq:hotspot}
Q_{\rm GS} = A c \frac{B_{\rm eq}^2}{8\pi} g =   2 \pi R^2 c \frac{B_{\rm eq}^2}{8\pi} g ,
\end{equation}
where $g$ is a function of various parameters of the flow causing the emission, but for simplicity's sake, we assumed $g \approx 2$ based on the AGN case.\footnote{The term $g$ is a complicated function of the spectral index of the radio emission $\alpha$, the velocity of the flow causing the emission, its density, the lepton and proton number densities, and of the magnetic field in the flow. Since we cannot easily estimate all of these quantities without making wild assumptions, we decided to adopt the same value used in the case of AGNs.} In Eq. \ref{eq:hotspot}, $A$ is the area of the  hot spot surface, which we assumed is a circular surface with radius $R \approx 7.6 \times 10^{18}~{\rm cm}$ based on the area of the northern feature, and B$_{eq}$ is the equipartition magnetic field, which from Sect. \ref{sec:cylinder} is approximately $22~{\rm \mu G}$. 
We thus have 
$ Q_{\rm GS} \sim 1.7 \times 10^{36}~{\rm erg~s}^{-1} $.
We note that this value is a lower limit since departures from equipartition will increase it. Still, it is consistent with the lower limit to the power estimate given in Sect. \ref{sec:calorimetry}.

\begin{table*}
\caption{Parameters measured and inferred in this work.}              
\label{tab:BHpar}      
\centering                          
\begin{tabular}{c c c}        

\hline                        
\multicolumn{3}{c}{\textbf{Source-related parameters}}  \\ 
\hline   

Distance$^{m}$                                  &       \multicolumn{2}{c}{9.4$\pm$0.7~kpc}                                     \\
Jet inclination angle$^{m} i $                  &       \multicolumn{2}{c}{60$\pm$5 deg}                                                \\
Jet opening angle$^{m} $$\theta$                &       \multicolumn{2}{c}{1-10 deg}                                                    \\

\hline                       

\multicolumn{3}{c}{\textbf{Thermal Bremsstrahlung region (IRAS 19132+1035) -  \textit{the IRAS region} }}                               \\ 
\hline   
Diameter$^{m}$                                  &       36 arcsec                   &   1.6 pc                                           \\
Volume$^{i}$                                    &       \multicolumn{2}{c}{6.6$\times$10$^55$ cm$^3$}                                     \\
Integrated flux density$^{m}$                   &       \multicolumn{2}{c}{60$\pm$6 mJy}                                                \\
Electron temperature$^{a}$                      &       \multicolumn{2}{c}{10$^4$ - 3$\times$10$^6$ K}                                           \\
Shock-compressed gas electron density$^{i}$     &       \multicolumn{2}{c}{395 - 630 particles/cm$^3$}                                  \\   
ISM gas density$^{i}$                           &       \multicolumn{2}{c}{100 - 160 particles/cm$^3$}                                    \\   
Shock front velocity                            &       \multicolumn{2}{c}{21 - 363 Km\,s$^{-1}$}                                               \\
Region gas pressure$^{i}$ (ideal gas)           &       \multicolumn{2}{c}{2.5$\times$10$^{-10}$ - 7.7$\times$10$^{-8}$ [erg/cm3]}        \\

\hline
\multicolumn{3}{c}{\textbf{Non-thermal emission region (cylindrical hot spot) - \textit{the northern feature} }}                        \\ 
\hline 
 
Flux density$^{m}$                              &       \multicolumn{2}{c}{5.2$\pm$0.5 mJy}                                                \\
Luminosity                                      &       \multicolumn{2}{c}{2$\times 10^{+29}$ erg\,s$^{-1}$}                                    \\
Radius$^{m}$                                    &        3.9  arcsec                & 0.18 pc                                           \\
Length$^{m}$                                    &        17.4 arcsec                & 0.9 pc                                            \\
Volume$^{i}$                                    &       \multicolumn{2}{c}{2.6$\times$10$^{54}$ cm$^3$}                                   \\
Cylinder pressure$^{i}$  (minimum energy)       &       \multicolumn{2}{c}{$>$ 1.5$\times$10$^{-11}$ erg/cm$^{3}$ }                     \\
Cylinder magnetic density B$_{min}^{i}$  (minimum energy)       &       \multicolumn{2}{c}{$>$ 2.18$\times$10$^{-5}$ Gauss}                              \\

\hline
\multicolumn{3}{c}{\textbf{The lobe and the bow shock structure} }                                                               \\ 
\hline   

Bow shock structure flux density$^{m}$                    &       \multicolumn{2}{c}{0.15$\pm$0.01 mJy}                                           \\
Bow shock distance from binary$^{m}$ (cavity major axis)             &        17 arcmin    &    46 pc                                                        \\
Lobe minor axis$^{m}$                                &        10 arcmin     &    27 pc                                                        \\
Bow shock structure thickness$^{m}$                                 &        60 arcsec    &    2.7 pc                                                       \\
Lobe volume$^{i}$                                    &       \multicolumn{2}{c}{6$\times$10$^{59}$cm$^3$}                                    \\
Lobe pressure$^{i}$  (equi-partition)           &       \multicolumn{2}{c}{$>$ 1.3$\times$10$^{-12}$ erg/cm$^{3}$ }                     \\
Lobe magnetic field$^{i}$ (synchrotron emission)&       \multicolumn{2}{c}{$>$ 6.4$\times$10$^{-6}$ Gauss}                              \\

\hline
\multicolumn{3}{c}{\textbf{Jet energetics}}  \\
\hline
Jet age$^{i}$                                   &       \multicolumn{2}{c}{0.09 - 0.22 Myr}                                             \\
Jet pressure$^{i}$  (self-similar model)     &       \multicolumn{2}{c}{2.8$\times$10$^{-11}$ -- 7.8$\times$10$^{-10}$  [erg/cm3]}     \\
One-sided power transferred (self-similar)$^{i}$                         &       \multicolumn{2}{c}{3.3$\times$10$^{37}$ - 1.5$\times$10$^{39}$ erg\,s$^{-1}$}                     \\
One-sided power transferred (Enthalpy)$^{i}$                         &       \multicolumn{2}{c}{> 1.2$\times$10$^{37}$ erg\,s$^{-1}$}                     \\
One-sided power transferred (Hot spot)$^{i}$                         &       \multicolumn{2}{c}{> 1.7$\times$10$^{36}$ erg\,s$^{-1}$}                     \\
Pseudo-constant C$_1^{i}$                       &       \multicolumn{2}{c}{3.3-9}                           \\

\hline                                   

\end{tabular}
\tablefoot{The table including the parameters measured (denoted with $^{m}$) or inferred (denoted with $^{i}$) in this work. Parameters have been divided based on the region they refer to. Where lengths are given in angular units and physical units, we intend the former as apparent, and the latter as corrected for projection effects, when appropriate.}
\end{table*}

\section{Discussion}\label{sec:discussion}

In the MeerKAT observations of the BH X-ray binary GRS 1915+105 we found an extended arch-like structure located to the south-east of the system, extending at either side of the bright source IRAS 19132+1035. IRAS 19132+1035 is located at a distance consistent with that of GRS 1915+105, and was identified as a potential jet-ISM interaction zone attributed to GRS 1915+105 \citep{Rodriguez1998,Chaty2001}. The association of IRAS 19132+1035 with GRS 1915+105 was recently confirmed by \cite{Tetarenko2018}, who mapped the molecular line emission across IRAS 19132+1035, and found evidence that supports the presence of a jet-ISM interaction at this site. As already noticed by \cite{Rodriguez1998}, to the north of IRAS 19132+1035 a linear feature with non-thermal spectrum is observable (referred to as \textit{the northern feature} in this paper). Such a feature has been interpreted as associated with the large-scale jet, an association that was confirmed in \cite{Tetarenko2018} based on molecular line features which indicate that the jet is pushing into dense gas.

\subsection{Bow shock structure comparison with Cyg X-1}

Based on the similarity in morphology to the bow shock structure observed near Cyg X-1 \citep{Gallo2005}, and on the apparent association with IRAS 19132+1035, we postulated that the arched structure near GRS 1915+105 formed as the result of the interaction of the jets with the ISM. In particular, we interpret the structure as the result of the jets from the binary ending in strong shocks at the location where they interact with the surrounding medium and blow a lobe into the ISM. The gas surrounding the cavity is thus compressed, and this results in a bow shock forming at the jet end. 

Compared to the case of Cyg X-1, the jet-induced structure near GRS 1915+105 shows greater complexity, featuring a dim bow shock structure, previously unknown and similar to that discovered near Cyg X-1 \citep{Gallo2005},  a bright, flat-spectrum extended emission region (the IRAS region), and a steep-spectrum linear structure adjacent to it, pointing towards the position of GRS 1915+105. 
In the jet-related scenario that we depicted, the IRAS region corresponds to compressed and shock-heated ISM in front of the jet, and the northern feature corresponds to the strong shock that originates at the end of the postulated large-scale jet, which in AGNs would be the hot spot. The bow shock structure corresponds to the shock-compressed material at the edges of a jet-blown cavity and was not found in previous searches for similar features \citep{Rodriguez1998, Heinz2003}. 
Under the assumption that the structure near Cyg X-1 and GRS 1915+105 are both jet-induced, the main difference between the two - i.e. the absence of a \textit{hot spot} in Cyg X-1 - could be explained in terms of particle acceleration at the jet end, which is efficient in GRS 1915+105, and inefficient or possibly absent in the case in Cyg X-1. It is also worth noting that the jet of GRS 1915+105 terminates at a molecular cloud, and hence it is presumably ramming into a denser medium than in the case of Cyg X-1, where the jet seems to interact with a less dense molecular cloud. 

\subsection{Properties of the interstellar medium}

Beside IRAS 19132+1035, also a candidate jet-ISM interaction spot with the counter-jet was identified to the north-west of GRS 1915+105 (IRAS 19124+1106). However, we do not detect any radio lobe near this region. If IRAS 19124+1106 is indeed an interaction site with the jet, then a possible bow shock structure (and lobe) would overlay a bright extended HII region (which IRAS 19124+1106 may or may not be part of). In either case, it would be hard to identify a possible bow shock structure, assuming its brightness would be similar to the structure at the south-east of GRS 1915+105. 

From the flux density of the thermal emission from the shock compressed gas we inferred the density of the ISM in the region of the shock. We find a particle density of the order 100 particles per cm$^3$, which is relatively high, but not excessively so given that this region of the sky is located on the Galactic plane, and is rich in dense star-forming regions. The density we found supports the hypothesis that the bow shock formed because of the interaction of the jets from GRS 1915+105 with a large HII region, similar to what was proposed for the case of Cyg X-1. This would help reconcile the apparent contrast between the gas density we derive, and the low values predicted by, for example, \citealt{Heinz2002, Heinz2003, Carotenuto2024}, who suggest that X-ray binaries may be located in low-density \textit{cavities} in the ISM.\\

\subsection{Jet properties}

Based on our conservative estimate of the temperature of the emitting gas within the IRAS region, we inferred the jet age and the jet time-averaged energy transport rate. This was under the assumption that the IRAS region is dominated by Bremsstrahlung emission indicated by its flat spectrum. We estimate that the jet that generated the bow shock must be between 0.09 and 0.22\,Myr old, with an upper limit of $\approx$0.4 (95th percentile; see Sect. \ref{sec:jetage}).  
If our assumptions are correct, the time required for the jets from GRS 1915+105 to slice across the IRAS region as the system moves through the Galaxy should be longer than the jet age itself. On the other hand, if the projected motion of GRS 1915+105 implies that its jet crossed the IRAS region faster than the age of the jet calculated using the temperature of the gas, the motion of GRS 1915+105 will give a stronger constraint on the jet age suggesting that this might be a better indicator of jet age than the analysis presented in this work.  
The projected proper motion of GRS 1915+105 on the plane of the sky is oriented approximately north-east to south-west (top left to bottom right in the image in Fig. \ref{fig:1915_field}), i.e. essentially orthogonal to the direction of its jets \citep{Dhawan2007}. Thus, we assume that the time GRS 1915+105 will need to move across the IRAS region is similar to the time the jets from GRS 1915+105 could have interacted with the IRAS region. Estimates of the potential kick velocity of GRS 1915+105 show that the BH in the system was born with a direct collapse, and thus the system's peculiar velocity is comparable to the local standard of rest \citep[][]{Atri2019,Reid2023}. Considering these two facts, and using the most recently reported non-circular velocity of GRS 1915+105 of 20\,km\,s$^{-1}$, we estimate that the time GRS 1915+105 (along with its jets) will take to move across the IRAS region is approximately 0.9 Myr, assuming a width for the IRAS region of 7.2\,arcmin. As expected, this time period is larger than the jet age, from which we can estimate a much more stringent lower limit on the jet power.

The one-sided time-averaged energy transport rate ranges between 3.3$\times$10$^{37}$ and 1.5$\times$10$^{39}$ erg\,s$^{-1}$, which we derived using the self-similar model proposed by \cite{Kaiser2004}. This estimate is confirmed by independent though simplified calculations based on arguments usually applied to AGNs. The first method is based on the estimate of the enthalpy of the regions powered by the jets, and the second yields the jet power under the assumption of equipartition applied to the jet \textit{hot spot}. In both cases, we obtain a strict lower limit to the jet power of 1.7$\times$10$^{36}$erg\,s$^{-1}$. 
The jet power estimated above exceeds by at least three orders of magnitude the radiative luminosity\footnote{The monochromatic radiative luminosity of the bow shock structure is $\sim 7 \times 10^{23}$ erg s$^{-1}$ Hz$^{-1}$ at 1.28 GHz. The bolometric radiative luminosity is estimated integrating between 10$^7$ and 10$^{11}$ Hz and assuming a spectral slope $\alpha$ = 0.7.} of the nebula, $\sim$1$\times$10$^{34}$ erg\,s$^{-1}$. This substantial difference indicates that the jet supplies far more energy than is required to power the observed radiation from the nebula. The excess power inferred from the jet's energy budget is likely deposited in ongoing particle acceleration and other processes that drive the physical evolution of the nebula. A similar situation is observed in the case of Cyg X-1 \citep[][submitted]{Atri2025}, where the bow shock radiative luminosity ($\sim$3$\times$ 10$^{31}$ erg s$^{-1}$ Hz$^{-1}$) is relatively low if compared to the jet power (between $\sim$1$\times$10$^{35}$\,ergs s${^{-1}}$-- 1$\times$10$^{38}$\,ergs s${^{-1}}$).

Our estimates of the time-averaged jet power differ quite significantly from the estimate derived by \cite{Tetarenko2018} ($\sim10^{32}$erg\,s$^{-1}$), which was also obtained following \cite{Kaiser2004} and arguments similar to those we adopted in this paper. The difference in the estimate can be ascribed to slightly different assumptions on the jet properties and on the composition of the IRAS region, and in particular (i) the jet opening angle (0-4 degrees in \cite{Tetarenko2018} versus 1-10 degrees in this work), (ii) the density within the IRAS region (4000 particles/cm$^3$ versus $\sim$ 80-180 particles/cm$^3$), (iii) the jet front expansion velocity (1 km\,s$^{-1}$ versus 20-360 km\,s$^{-1}$ ). 
The assumption on the jet opening angle is rather speculative and does not affect significantly the lower limit to the time-averaged energy transport rate (set by the lower limit on the jet opening angle). The particle density in \cite{Tetarenko2018} is estimated based on the CO density maps from ALMA data, which are used to estimate the HII column density. This value is necessarily associated with the cold gas in the region, which in this work is postulated to fill only 50\% of the IRAS region, while the remaining volume is filled with hot gas (see Sect. \ref{sec:iras}), which is generating the radio continuum emission considered here. The highest impact (a difference of the order $10^4$), is given by the shock front propagation velocity, which in \cite{Tetarenko2018} is derived based on the equivalent width of the SiO lines, also arising from the cold gas, and thus not necessarily correct for the hot gas in the region.

\subsection{The IRAS region}

The existence of multi-phase gas with temperatures of $\sim10-100~{\rm K}$ and $> 10^4~{\rm K}$, as already suggested by \cite{Kaiser2004} is required by the simultaneous presence of both the flat radio continuum and the lines emission. The ISM is inherently multi-phase \citep{tielens2005} -- but how does this cold phase survive being shocked? This apparent conundrum can be resolved by considering the different shock propagation velocities through the two gas phases. If the shock strikes a multi-phase ISM region in pressure equilibrium, then the density contrast $\chi$ between the molecular clouds and warm phase is the inverse of their temperature ratio, implying $\chi \sim 10^2-10^3$. The question of how long the molecular clouds survive is critical and corresponds to the well-studied `cloud-crushing' problem, where the time for a shock to pass through an overdense cloud (in our case the cold, denser molecular gas) is obtained from ram pressure balance \citep{klein1994} and given by
\begin{equation}
    \tau_{\rm cc} = \chi^{1/2} {R_c}/{v_s},
\end{equation}
where $v_s$ is the shock velocity in the underdense, warm region, and $R_c$ is the radius of a dense cloud. Hence, we argue that the molecular line emission is observed because (i) the clouds have not had sufficient time to be shocked and disrupted (perhaps implying that $R_c$ is fairly significant compared to the width of IRAS 19132+1035) and/or (ii) the lower shock velocity within the molecular clouds means the post-shock temperature is much lower than that in the warm phase. The basic scenario is that of a shock passing quickly through the warm phase and leaving behind embedded dense molecular clouds that are disrupted in a few cloud-crushing times. This behaviour is essentially that depicted in \cite{mckee1977} (Fig.~2), and also occurs in simulations of the interaction of AGN jets with a two-phase ISM \citep{mukherjee2018}.


The morphology of the IRAS region in radio and in the sub-millimetre band and the presence of cold gas and molecular lines that can be ascribed to young star activity prompted \cite{Tetarenko2018} to evaluate what influence star formation had on the region and consider the possibility that star formation was triggered by the jet itself interacting with the environment. While the thermal component of the radio continuum could be explained in terms of the activity of a young star cluster, we argue that the discovery of a bow shock marking the presence of a cavity inflated by the jets from GRS 1915+105 suggests that the jet's impact on the region must have been significant. Hence, the star activity might not have been the dominant factor in shaping the IRAS region. 
As discussed in \cite{Tetarenko2018} and other authors \citep[see e.g. ][]{Podsiadlowski2003}, GRS 1915+105 is old enough that the jet activity might have indeed triggered star formation. The transient nature of GRS 1915+105 and the age of the jet estimated above imply that the bow shock structure we observe today may only be the last one formed by the jet, and much older activity might have occurred. 
We note in particular the presence of arched structures to the south-east of the IRAS region, following the same orientation of the bow shock structure. Such structures may be related to older bow shocks formed in previous bursts of jet activity, which expanded further out with respect to the IRAS region, and may be between 0.1 and 0.5 Myr old. 

\subsection{Implications}

In the scenario we have depicted, the energy rate transported by the large-scale jet that induced the formation of the bow shock is comparable with the average energy typical of the small-scale jets observed during outbursts. The average energy rate is estimated to be between 10$^{37}$ and 10$^{42}$ erg\,s$^{-1}$ for both the quasi-continuous jets typical of the harder states, and the transient discrete jet observed at sub-arcseconds scales during the intermediate states \citep{Fender1999, Fender2000, Heinz2005}. Especially the latter is believed to be launched on short time scales, which may imply that the dark jet either transports similar amounts of energy as the transient jets and is active over similarly short scales or it transports less energy but is active for longer times. In this regard, it might be possible that the presence of a hot spot in the bow shock structure near GRS 1915+105 - a feature that is absent in the case of Cyg X-1 - might be connected with the possible turning on and off of the jet activity, as opposed to the quasi-continuous jet from Cyg X-1 (Cyg X-1 also goes through occasional soft states). It is also interesting to notice that the minimum time-averaged energy transferred by GRS 1915+105 is at least an order of magnitude larger than what is estimated for the case of Cyg X-1 (1.4$\times$10$^{35}$\,ergs s$^{-1}<$ Q$_{jet}^{a}<$ 1.4$\times$10$^{36}$\,ergs s$^{-1}$, \citealt[][submitted]{Atri2025}, \citealt{Gallo2005}). 

Since GRS 1915+105 is a transient source and not a persistent one as Cyg X-1, it is hard to correct the time-averaged energy rate transported by the dark jet we estimate above to account for the varying activity period of the system. The estimated age of the bow shock near GRS 1915+105 is at least 80~kyr, which indicates that the structure must have formed much earlier than during the current outburst (started in 1995), that is, either during previous outbursts (undetected, because they were precedent to the birth of X-ray astronomy) or during quiescence.
In the former case, the time-averaged transferred energy in GRS 1915+105 may be up to two orders of magnitude larger than what we estimated, as it must take into account the fact that an outburst that may last around 30 years is between 25 and 100 times shorter than the age of the jet that generated the bow shock.   
If, instead, the bow shock formed during quiescence, then it must be connected with the radio-faint continuous jets observed in these phases, when the accretion energy dissipated as X-ray is negligible \citep{Fender2004}. If this is the case, our estimate of the time-averaged transferred energy would be more accurate. 

\section{Conclusions}\label{sec: conclusions}

We have reported the discovery of a bow shock structure towards GRS 1915+105 observed in recent MeerKAT observations at 1.28 GHz. Assuming the structure is associated with GRS 1915+105, we attributed it to the presence of a large-scale dark jet that is interacting with the local environment several parsecs away from the launching site.
This finding provides further direct evidence of jet-ISM interactions, a crucial ingredient of BH feedback, the understanding of which has been based so far mainly on AGN studies. 
Our analysis reveals that the energy dissipated by the jets in this system is comparable to the energy liberated via accretion, indicating that stellar-mass BHs can transfer substantial amounts of energy back into their environments through jet activity. 

Our findings have important implications for the study of X-ray binaries as a population. The detection of jet-induced large-scale structures first near Cyg X-1 (a persistent system) and now in the vicinity of GRS 1915+105 (a very variable and transient system, though it has been in outburst for 30 years at the time of writing) demonstrates that the impact of jets from Galactic compact objects may extend far beyond their immediate vicinity, potentially influencing the ISM by enriching it with matter, re-heating it with energy, and even potentially triggering star formation, hence significantly contributing to the broader ecology of the Galaxy. 

In the broader context of accreting BHs at all scales, our results indicate that jet-induced feedback mechanisms are not exclusive to SMBHs but are also relevant for their stellar-mass counterparts and not necessarily only those hosted in X-ray binary systems. This expands our understanding of how BHs of different masses impact their cosmic environments, from local star formation to the evolution of galaxies.

\bigskip 
\bigskip



\bigskip

\begin{acknowledgements}

The authors would like to thank Dave Russell, David Williams-Baldwin, Cristina Baglio, Fraser Cowie, and Andrew Hughes for useful feedback on this work, and Evangelia Tremou and Joe Bright for scheduling the observations that led to the results reported in this paper.\\ 

SEM acknowledges support from the INAF Fundamental Research Grant (2022) EJECTA.\\

The MeerKAT telescope is operated by the South African Radio Astronomy Observatory, which is a facility of the National Research Foundation, an agency of the Department of Science and Innovation.

We acknowledge the use of the Ilifu cloud computing facility – www.ilifu.ac.za, a partnership between the University of Cape Town, the University of the Western Cape, Stellenbosch University, Sol Plaatje University, and the Cape Peninsula University of Technology. The Ilifu facility is supported by contributions from the Inter-University Institute for Data Intensive Astronomy (IDIA – a partnership between the University of Cape Town, the University of Pretoria and the University of the Western Cape), the Computational Biology division at UCT and the Data Intensive Research Initiative of South Africa (DIRISA).\\

PA is supported by the WISE fellowship program, which is financed by NWO.
JvdE acknowledges a Warwick Astrophysics prize post-doctoral
fellowship made possible thanks to a generous philanthropic do-
nation.

This work made use of the CARTA (Cube Analysis and Rendering Tool for Astronomy) software (DOI 10.5281/zenodo.3377984 – \url{https://cartavis.github.io)}.

\end{acknowledgements}

\section*{Data availability}

The un-calibrated MeerKAT visibility data presented in this paper are publicly available in the archive of the South African Radio Astronomy Observatory at https://archive.sarao.ac.za. The capture blocks considered in this work are listed in Appendix \ref{App:App1}.
The Continuum MeerKAT observations were taken as part of the ThunderKAT Large Survey Program, project code SCI-20180516-PW-01. 
The python source code used to perform the calculations is available online on GitHub \url{https://github.com/saramotta/DarkJets}. 

%
%

\bibliographystyle{aa.bst}
\bibliography{biblio} 

\newpage
\begin{appendix} 

\section{Additional information}\label{App:App1}

\subsection{MeerKAT observations}\label{sec:app_obs}

\begin{table}[h!]
\centering
\begin{tabular}{|c|c|c|}
\hline
1544268663 & 1593213962 & 1605454262 \\
1563825653 & 1593890159 & 1606051862 \\
1564248236 & 1594580523 & 1609148467 \\
1564947058 & 1595185558 & 1610180483 \\
1567871158 & 1595623962 & 1610782322 \\
1583118973 & 1597515368 & 1611982081 \\
1585883785 & 1598130064 & 1614495843 \\
1587176616 & 1598801721 & 1615005089 \\
1587786356 & 1599928833 & 1615601592 \\
1588375604 & 1600623068 & 1616209876 \\
1588992355 & 1601142995 & 1617595273 \\
1589683557 & 1601731867 & 1618018273 \\
1590975056 & 1602343380 & 1650081167 \\
1592167259 & 1602938168 & 1650081167 \\
1592527255 & 1603632965 & 1686009070 \\ 
\hline
\end{tabular}
\vspace{0.5cm}
\caption{MeerKAT observations used in this work, available on the SAREO archive. The codes reported correspond to capture block IDs. }
\end{table}\label{tab:Ids}

\subsection{The IRAS region}\label{sec:App_IRAS}

The monochromatic emissivity $\epsilon_\nu$ of ionised hydrogen due to thermal bremsstrahlung is given by \citep{Longair1994}: 
\begin{equation}
    \epsilon_\nu = \frac{L_\nu}{V} = C_{\rm radio} g(\nu, T) \frac{n_e^2}{\sqrt{T}} \exp{(\frac{h\nu}{k_b T})} ~~ erg~s^{-1}~cm^{-3}~Hz^{-1}
\end{equation}\label{eq:emissivity}

\noindent where $ C_{\rm radio} = 6.8\times 10^{-38} {\rm erg~s}^{-1}~{\rm cm}^{-3}~{\rm Hz}^{-1} $, and the Gaunt factor $g(\nu)$ is given by: 
\begin{equation}
g(\nu, T) \approx \frac{\sqrt{3}}{2 \pi}\left[\ln \left(\frac{128 \epsilon_0^2 k^3 T^3}{m_e e^4 \nu^2 Z^2}\right)-\gamma^2\right]
\end{equation}\label{eq:gaunt}

\noindent From Eq. \ref{eq:emissivity} we can derive the electron density $n_e$:

\begin{equation}
n_e = \sqrt{\frac{\epsilon_\nu}{g(\nu, T) \sqrt{T} C_{\rm radio} \exp{(\frac{h\nu}{k_b T})} } }
\end{equation}

\noindent where we assumed that the number of electrons $n_e$ equals the number of protons $n$ (i.e. pure hydrogen gas), and that the density and temperature are uniform within the volume $V$ considered, i.e. in the case of the IRAS region a sphere with apparent diameter 36 arcsec.

\subsection{The northern feature }\label{sec:App_northern}

To model the non-thermal region, we assume a radio spectrum with a spectral index $\alpha$ between upper and lower frequencies $\nu_1, \nu_2$. We used the integrated radio luminosity between these limits (and not the specific radio luminosity at a certain frequency\footnote{We note that the equivalent formula used in \cite{Longair1994} gives the 'specific' luminosity in units of erg s$^{-1}$ Hz$^{-1}$ $L_{\nu}$, whereas the integrated luminosity used here is erg s$^{-1}$ and is therefore a much larger quantity.}) given by
\begin{equation}
L=4 \pi D^2 \int_{\nu_1}^{\nu_2} F_{\nu} d\nu = 4 \pi D^2 F_{\nu_2} \nu_2^{-\alpha} \left( \frac{\nu_2^{\alpha+1} - \nu_1^{\alpha+1}}{\alpha+1} \right) ,
\end{equation}
where we have assumed $\alpha = (1-p)/2$ = -0.75 \citep{Mirabel1994}.

The energy density in magnetic fields is $B^2 / 8\pi$ (in cgs units); therefore, for a volume $V$, the total energy in the magnetic field is
\begin{equation}\label{eq:EB1A}
E_B = \frac{B^2}{8 \pi} f V, 
\end{equation}
where $f$ is a filling factor, which represents the fraction of the physical source's volume which is actually occupied by the magnetic field and relativistic particles. In our case $f$ = 1.

Given a population of electrons with a power law distribution with index $-p$ between energy limits $E_1$ and $E_2$, the total energy in these electrons is given by
\begin{equation}
E_e = c_2^{-1} L B^{-2} \frac{(p-3)}{(p-2)} \frac{(E_1^{2-p}-E_2^{2-p})}{(E_1^{3-p}-E_2^{3-p})}.
\end{equation}
For a given electron energy and magnetic field, we can characterise the synchrotron emission as coming from a single frequency $\nu = c_1 B E^2$, that is, $E = \nu^{1/2} c_1^{-1/2} B^{-1/2}$. So, we can write
\begin{equation}\label{eq:EeA}
    E_e = c_2^{-1} c_1^{1/2} \tilde{c}(p, \nu_1, \nu_2) B^{-3/2} L = c_{12}(p,\nu_1,\nu_2) B^{-3/2} L,
\end{equation}
where
\begin{equation}
c_{12} = c_2^{-1} c_1^{1/2} \tilde{c}(p, \nu_1, \nu_2)
\end{equation}
and
\begin{equation}
\tilde{c}(p, \nu_1, \nu_2) = \frac{(p-3)}{(p-2)} \frac{\nu_1^{(2-p)/2}-\nu_2^{(2-p)/2}}{\nu_1^{(3-p)/2}-\nu_2^{(3-p)/2}}.
\end{equation}

\noindent Since $E_B \propto B^2$, the total energy is given by
\begin{equation}\label{eq:EtotA}
E_{e+B} = E_e + E_B =  const. B^{-3/2} + const. B^2,
\end{equation}
and thus there must be a minimum of the total energy as a function of $B$ if we assume that there are no other contributions to the energy budget. Such a minimum occurs at the equipartition of the energy, that is, when $E_e$ and $E_B$ are equal (which can be calculated by differentiating Eq. \ref{eq:EtotA} with respect to $B$):

\begin{equation}\label{eq:EB2A}
    E_B = \frac34 \eta E_e,
\end{equation}
where $\eta$ indicates how much energy is stored in the protons which accompany the electrons (in the case of a normal baryonic plasma). For the minimum energy condition $\eta = 1$ is generally assumed.
Using Eqs. \ref{eq:EB1A}, \ref{eq:EeA}, and \ref{eq:EB2A}, we can express the equipartition magnetic field as
\begin{equation}\label{eq:BeqA}
    B_{\rm eq} = \left(6 \pi \frac{\eta}{f} c_{12} \frac{L}{V}\right)^{2/7}.
\end{equation}
For minimum energy conditions, the total pressure is then given by
\begin{equation}\label{eq:peqA}
    p = \frac{7}{9} \frac{B^2_{min}}{8\pi}(k+1),
\end{equation}
with $k$ equal the ratio of the internal energy stored in any particles that do not contribute to the synchrotron emission to the sum of the energy in the magnetic field and in the relativistic electrons.

\subsubsection{The $\bar{C1}$ dimensionless constant}\label{sec:C1}

The term $\bar{C1}$ is a dimensionless constant that depends on the thermo-dynamical properties of the jet material, and on the axial ratio R$_{ax}$ of the lobe inflated by the jet, which equals the length of the lobe divided by its width if the lobe is assumed to be cylindrical. 
Such a constant can be calculated by making assumptions on the jet properties and on the environment, which in our case is treated as uniform in density (i.e. the density profile index $\beta$ equals 0). The constant depends heavily on the jet opening angle $\theta$, as well as on the adiabatic indices of the material in the lobe cavity ($\Gamma_c$), in the external medium ($\Gamma_x$), and within the jet ($\Gamma_j$). 

The explicit form of $\bar{C1}$ is the following:
\begin{equation}
\bar{C_1}=\left(\frac{\bar{C_2}}{\bar{C_3}\theta^2}\frac{(\Gamma_x+1)(\Gamma_c-1)(5-\beta)^3}{18\left[9\{\Gamma_c+(\Gamma_c-1)\frac{\bar{C_2}}{4\theta^2}\}-4-\beta\right]}\right)^{\frac{1}{(5-\beta)}},
\end{equation}
with 
\begin{equation}
\bar{C_2}=\left(\frac{(\Gamma_c-1)(\Gamma_j-1)}{4\Gamma_c}+1\right)^{\frac{\Gamma_c}{(\Gamma_c-1)}}\frac{(\Gamma_j+1)}{(\Gamma_j-1)}
\end{equation}
and 
\begin{equation}
    \bar{C_3}=\frac{\pi}{4R_{\rm ax}^2},
\end{equation}
where R$_{ax} = \sqrt{\frac{1}{4}\frac{\bar{C_2}}{\theta^2}}$. 

\end{appendix}

\end{document}